# Nanosized patterns as reference structures for macroscopic transport properties and vortex phases in YBCO films

E. Mezzetti, A. Chiodoni, R. Gerbaldo, G. Ghigo, L. Gozzelino, B. Minetti

*INFM - U.d.R Torino-Politecnico; INFN - Sez. Torino*
*Politecnico di Torino, c.so Duca degli Abruzzi 24, 10129 Torino, Italy*

C. Camerlingo
*Istituto di Cibernetica del Consiglio Nazionale delle Ricerche*
*Via Toiano 6, 80072 Arco Felice, Italy*

C. Giannini
*Pa.S.T.I.S.–C.N.R.S.M., SS 7 Appia km.712, 72100 Brindisi, Italy*

This paper studies the striking correlation between nanosized structural patterns in YBCO films and macroscopic transport current. A nanosized network of parallel Josephson junctions laced by insulating dislocations is almost mimicking the grain boundary structural network. It contributes to the macroscopic properties and accounts for the strong intergranular pinning across the film in the intermediate temperature range. The correlation between the two networks enables to find out an outstanding scaling law in the ($J_c$,B) plane and to determine meaningful parameters concerning the matching between the vortex lattice and the intergranular defect lattice. Two asymptotic behaviors of the pinning force below the flux flow regime are checked: the corresponding vortex phases are clearly individuated.

HTSC films exhibit $J_c$ higher than single crystals. Moreover, in the intermediate temperature range films exhibit $J_c$ vs. log*B* curves with a characteristic plateau, absent in single crystals [1]. As a consequence, any insightful study of transport properties of films has to connect them to their particular microstructure as well as provide a model to interpret the plateau-like features.

The first striking difference between single crystals and films consists in the granularity of the latter (in this paper we do not consider the cell-scale granularity, exhibited also by single crystals and showing up in the stripes properties [2]). The effects of granularity on superconducting films has therefore been an "hot point" in HTSC investigation [3]. In fact, a lot of studies were performed on bicrystals of various kind, because these bicrystals can offer good model systems to obtain detailed information concerning the effect of grain boundary (GB) interfaces in films. It must however be emphasized that they offer a good model for what concerns granularity appearing at a mixed nanometric-micrometric scale (nanometric for single dislocations, micrometric for larger mismatch zones [4]), while only occasionally the impact of granularities appearing at just a gauge scale (it means at a few nanometer scale) can be guessed [5]. Besides an extreme care in this respect [6,7], the studies concerning low angle grain boundaries (GB) are particularly appropriate to the matter of films. It was found that this grain boundary consists of an array of edge dislocations that accommodate the lattice mismatch across the boundary [8]. The region between the edge dislocations can consist of relatively undisturbed lattice. This picture is easily compared, from a qualitative point of view, with the following framework. In good quality films the dislocation region behaves as a very small, "hidden" Josephson junction (JJ), separated from the contiguous parallel JJs by insulating larger dislocations [9, 10]. The insulating dislocations are likely located either where GBs cross each other or where the order parameter suppression, due to larger defects, hamper charge tunneling across short junctions. At the cross, an intrinsic columnar defect can be guessed. This picture is at the basis of a new model, which gave a quantitative account of the $J_c$ vs. *B* "anomalous" behavior of films. In Fig.1a (inset) we present a sketch of the hidden network of JJs, insulated by larger dislocations and interlacing the film, like the one we assume to be responsible for the transport across the film. An average 1D network the transport current could be driven across, is also outlined in bold.

In this letter we report a detailed comparison between the analysis of magnetic transport measurements, made by means of such model, and x-ray diffraction analysis on $YBa_2Cu_3O_{7-\delta}$ (YBCO) films. We show that the nanosize grain boundary network, broken as an array of small JJs, and a "structural" nanosized boundary network are strongly correlated. We show that they are the two aspects of a unique physical reality. The almost perfect coincidence between the two networks is striking, thought it must obviously taken into account that x-rays are probing the microstructural ordering, while magnetic transport measurements probe carriers tunneling and flux dynamics in a given range of temperature. A comparison between the two networks enables to characterize the matching between the vortex lattice and the pinning lattice lying along the GBs, by means of the experimental evaluation of crucial parameters.

The YBCO films were fabricated by dc sputtering from inverted cylindrical magnetron on $SrTiO_3$ (STO) substrates. The growth procedure was described elsewhere [10]. The texture nucleate on the $SrTiO_3$ substrate in the form of c-axis oriented islands. An island structure is clearly visible by AFM imaging: the surface morphology consists of islands about 100 nm in diameter [10]. As it will be shown, these islands contains further sub-grains, which can be detected by the x-ray analysis described below. Thus the boundaries between islands are non-uniform interfaces, modulated by the presence of the sub-grains, whose dimensions are one order of magnitude lower than the islands [11]. This does not affect the single grain orientation, with c-axis oriented along the normal of film surface and the a-axis oriented in plane along the cubic cell side of the substrate, as clearly indicated by x-ray φ-map measurements [12].

Critical current values have been extracted from susceptibility measurements by means of the procedure described in Ref.10.

The critical current of the array of parallel uniform short JJs of different lengths L, accounting for the particular shape of $J_c$ vs. B curves, can be written as

$$J_c(B) = J_c(0)\int_0^\infty dL\ p(L)\left|\frac{sin(\pi B\Lambda_0 L/\Phi_0)}{\pi B\Lambda_0 L/\Phi_0}\right| \qquad (1)$$

where $p(L)$ is the statistical length distribution of the junctions and $\Lambda_0$ is a very crucial field-dependent "magnetic thickness" [13], that in the present case is assumed to be proportional to the mean vortex distance $a_0$: $\Lambda_0 = \zeta a_0$ [9, 10]. $\zeta$ is a number of order unity, which in a bicrystal generally depends on bulk pinning and on the geometry of the grain boundary [9]. In the context of the paper $\zeta$ will assume a more definite role. If we choose

$$p(L) = \frac{\mu^\nu}{\Gamma(\nu)} L^{\nu-1} e^{-\mu L} \qquad (2)$$

the expression (1) assumes the form

$$J_c(B) = J_c(0)\,\frac{1}{\Gamma(\nu)\,q\sqrt{B}}\int_0^\infty x^{\nu-2}e^{-x}\left|sin(q\sqrt{B}x)\right|dx \qquad (3)$$

where $q = \pi\zeta\langle L\rangle/(\nu\,\Phi_0^{1/2})$, $\nu = (\langle L\rangle/\sigma_L)^2$, $\langle L\rangle$ is the mean value and $\sigma_L^2$ the variance of the distribution. The fit of the experimental data (an example is shown in Fig.1a) allows determining the two crucial parameter $\langle L\rangle\zeta$ and $\langle L\rangle/\sigma_L$. However it does not provide the independent determination of $\langle L\rangle$, $\zeta$ and $\sigma_L$. In Fig 1b the length distribution is plotted for seven tentative values of $\zeta$, near the values quoted in previous literature [9].

In order to determine all the physical parameters a comparison between the carrier transport properties and the structural properties of the host materials on a nanosize gauge scale is needed.

The samples were characterized by XRD measurements performed by using a Philips-1880 diffractometer equipped with a Cu target as x-ray source, a 1/30° collimation slit and a 0.1-mm receiving slit. Fig.1c shows the x-ray diffraction profile measured, in a coupled θ–2θ movement, on the YBCO film grown on top of the (001)-oriented STO substrate. The spectrum shows several peaks belonging to the (00l) family planes of the YBCO and STO crystal lattices. We can conclude that the YBCO film shows a good c-axis orientation.

A statistical model was used to analyze the X-ray data through a lineshape fitting procedure based on a Montecarlo approach. This model describes the broadening of the diffraction peaks as the consequence of the frequency distribution of the grain size [14]. The inset of Fig.1d shows the fitting results obtained from the analysis of the YBCO (007) peak indicated by the arrow in Fig.1c. The mean grain size value, as derived from the statistical analysis, is 19±3 nm in diameter, and the complete frequency distribution of the grain diameter (D) is shown in Fig.1d.

As it can be easily guessed by comparing Fig.1b and Fig.1d, the frequency distribution of the grain diameter, 'structural curve', can be very well fitted with the statistical distribution (2). On the basis of this correspondence, we made the hypothesis that the JJ length distribution p(L) in these optimally textured films is an ensemble representative of the structural grain boundary dislocation lattice. We then used the curve of Fig.1d, once fitted with the statistical distribution p(L), to obtain new fits of the experimental $J_c$ vs. B curves. *In this second analysis the only free parameter left is* $\zeta$. The results are shown in Fig.2, where two final fitting for T=28K and 67K are plotted and *the temperature dependence of the factor* $\zeta$ *is reported.*

A tentative scaling of the $J_c(B)$ curves can be tried by plotting the data in proper normalized units $J_c(B)/J_c(0)$ vs. $B/B_0$. In fact, in the present case the argument of the *sin* function in eq.(1) can be rewritten as $\pi(B/B_0)^{1/2}$, where $B_0 = \Phi_0/\zeta^2 L^2$. Fig.3 shows the outstanding scaling of the $J_c(B)$ curves obtained at different temperatures when $B_0 = \Phi_0/\zeta^2\langle L\rangle^2$ is considered. The corresponding scaling law for the accommodation field B*(T), defined as the field where $J_c$ reaches a given fraction of the zero-field value [10], results to be

$$B^*(T)\zeta^2(T)\langle L\rangle^2 = \text{cost.} \qquad (4)$$

To assess the vortex pinning by the grain boundary network we study the dependence of the pinning force on field, related to the GB distribution of Fig.1d. Since the pinning force is defined as $F_p = J_c B$, its asymptotic behavior can be calculated from (3) in the limit of $J_c(0)$ = cost. The asymptotic behavior can be easily assessed with a straightforward analytical calculation if we use for the critical current the equivalent analytical expression of Ref.10. It results that for low fields (B→0) $F_p$~B and for high fields ($B \rightarrow \infty$) $F_p \sim B^{1/2}$. In Fig.4 the two asymptotic behaviors are plotted as continuous curves, while the experimental data are represented as open symbols. The measurements, as expected by the $J_c(B)$ plateau-like feature, fit well to a straight line in a log-log plot with exponent 1 at low fields; at higher fields the slope 1/2 is reached. The two asymptotic curves cross each other at a crossover field $B_{co}$.

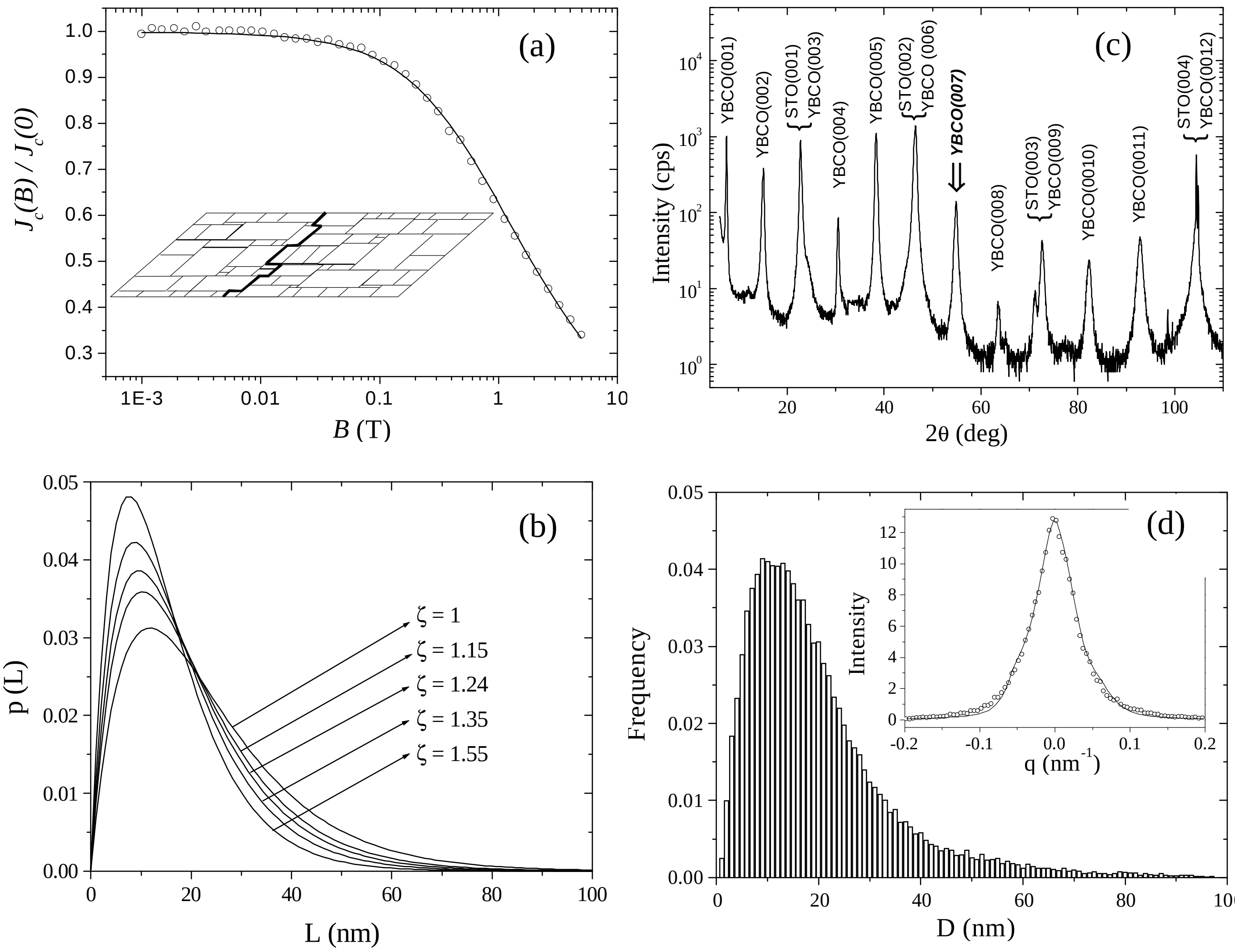

FIG.1. Critical current density vs. applied magnetic field, as deduced from the susceptibility measurement (a). $J_c$ experimental data (symbols) have been fitted (solid line) by means of equation (3). In the inset, a schematic view of the JJ network in the modulated interface between two ex-grown island is shown. The junction-length distributions resulting from the fit reported in (a) for different ζ values is shown in (b). Fig.1c shows the x-ray diffraction profile measured on the YBCO film grown on top of the (001)-oriented STO substrate. The inset of (d) shows the fitting results obtained from the analysis of the YBCO (007) peak, indicated by the arrow in (c). The complete frequency distribution of the grain diameter is shown in the main panel of Fig.1d.

In interpreting our data in a less quantitative, however appealing picture, we can consider the volume pinning force as proportional to the linear density of the pinned vortices, times the pinning force per unit length [15]. In the low field range the pinning force per unit length increases as the linear vortex density increases, because in this regime the entering vortices are supposed to find new assets and new matching conditions. This phenomenon occurs in less and less steady conditions until a saturation regime sets up where the pinning force per unit length becomes constant (slope 0.5) [15]. Obviously our "continuos" model is unable to account for the discontinuous entrance of the vortices into the network [16].

In conclusion we have found a striking correlation between nano-sized structural patterns and macroscopic transport current. A nano-sized JJ network of parallel junctions, contributing to the macroscopic properties is almost mimicking the grain boundaries structural network. It accounts for the strong intergranular pinning across the film.

In the investigated range of temperatures the comparison enables to assess experimentally the scale parameters in the ($J_c$,B) plane. The pinning force exhibits two asymptotic behaviors, easily referred to the network itself. Two corresponding asymptotic vortex phases are clearly individuated.

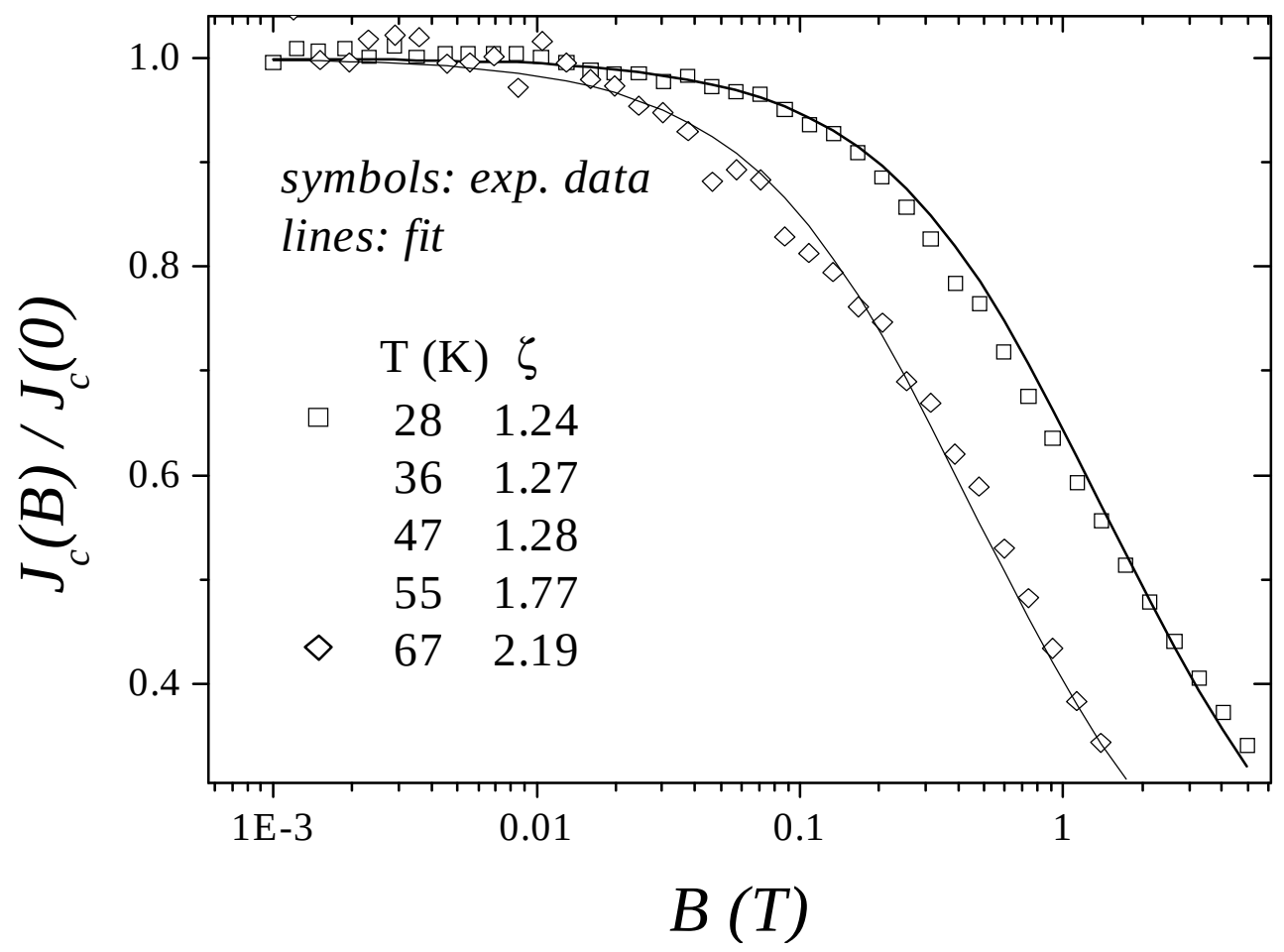

FIG.2. Critical current density data fitted by (3) and fitting parameter ζ as obtained on the basis of the grain-size distribution of Fig.1d.

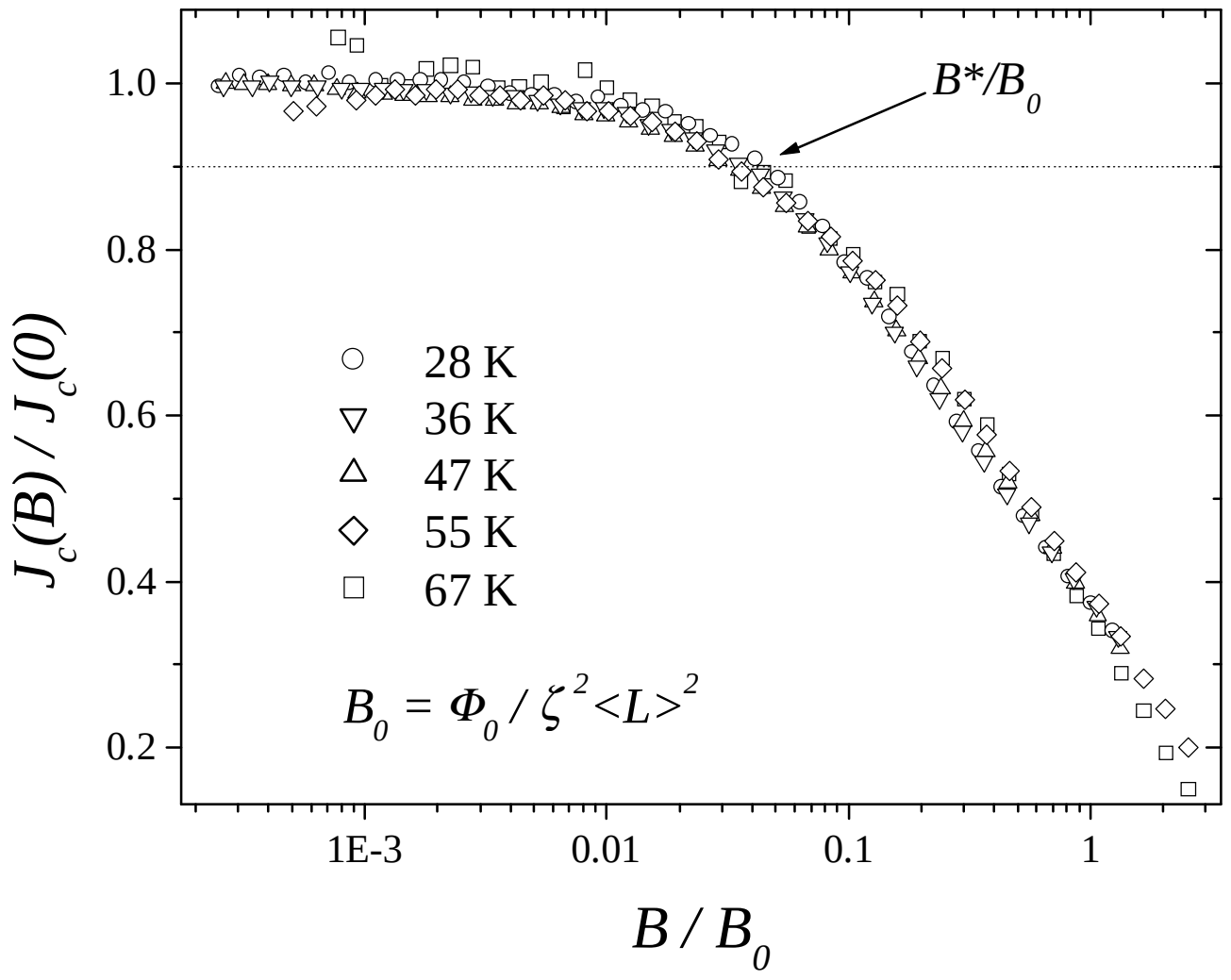

FIG.3. Scaling of the critical current density values at different temperatures.

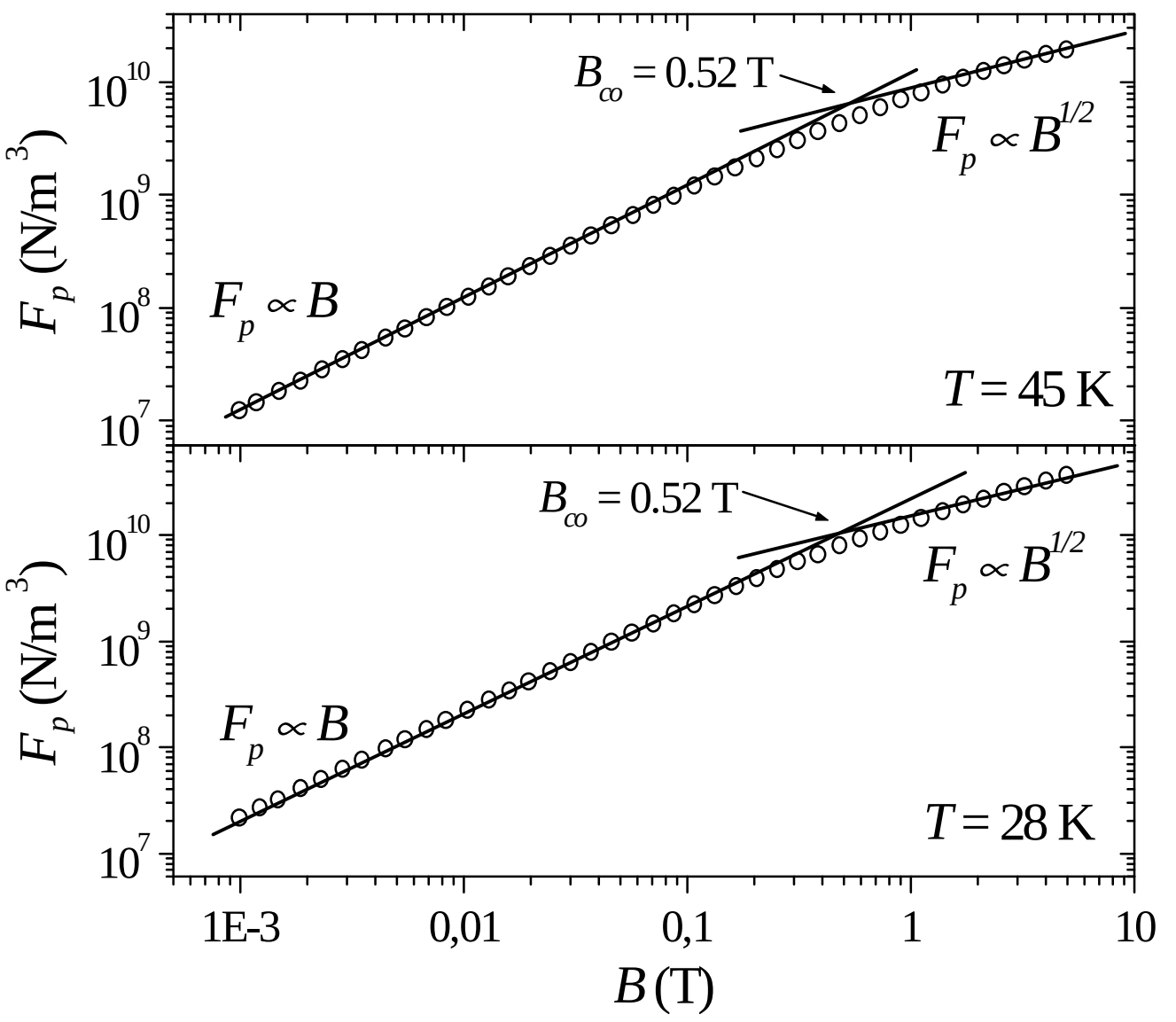

FIG.4. Pinning force as a function of the magnetic field at two different temperatures. Two distinct behaviours can be recognized, $F_p \propto B$ at low fields and $F_p \propto B^{1/2}$ at high fields. The crossover field $B_{co}$ is indicated with an arrow.

Work partially supported by a MURST COFIN98 program (Italy) and by ASI-ARS99 project. The authors wish to thank G.Costabile, C.Nappi, S.Martelli and L.Tapfer for helpful discussions.